\documentclass[sigconf]{acmart}
\renewcommand\footnotetextcopyrightpermission[1]{} 
\usepackage{graphicx}
\usepackage{amsmath}
\usepackage{booktabs}
\usepackage{amsfonts}
\usepackage{multirow}[c]
\usepackage{makecell}
\usepackage{subfig}
\usepackage{color}
\usepackage{bm}
\usepackage{epstopdf}
\usepackage{url}
\usepackage[cal=cm]{mathalfa}
\usepackage{balance}
\usepackage{threeparttable}
\usepackage{wrapfig}
\usepackage{tabularx}
\usepackage{array}
\usepackage{enumitem}
\usepackage{appendix}
\usepackage{tikz}
\usepackage{float} 
\usepackage{adjustbox}
\usepackage[linesnumbered,ruled,vlined]{algorithm2e}

\SetKwProg{Stage}{Stage}{}{} 

\setlist[itemize]{leftmargin=*}

\AtBeginDocument{}

\title{Unifying Generative Recall and Multi-Objective Ranking \\ in a Single Decoder-Only Sequence}

\setcopyright{acmcopyright}
\copyrightyear{2026}
\acmYear{2026}
\acmDOI{XXXXXXX}

\acmConference[Conference acronym 'XX]{Make sure to enter the correct
  conference title from your rights confirmation email}{June 03--05,
  2018}{Woodstock, NY}

\begin{document}

\author{Ruochen Yang}
\affiliation{
  \institution{Institute of Information Engineering, Chinese Academy of Sciences  \\ School of Cyber Security, University of Chinese Academy of Sciences}
  \city{Beijing}
  \country{China}
}
\email{yangruochen@iie.ac.cn}

\author{Shuang Wen}
\affiliation{
  \institution{Kuaishou Technology}
  \city{Beijing}
  \country{China} 
}
\email{wenshuang@kuaishou.com}

\author{Pengbo Xu}
\affiliation{
  \institution{Kuaishou Technology}
  \city{Beijing}
  \country{China} 
}
\email{xupengbo@kuaishou.com}

\author{Yusheng Huang}
\affiliation{
  \institution{Kuaishou Technology}
  \city{Beijing}
  \country{China} 
}
\email{huangyusheng@kuaishou.com}

\author{Jiangxia Cao}
\authornote{Corresponding author.}
\affiliation{
  \institution{Kuaishou Technology}
  \city{Beijing}
  \country{China} 
}
\email{caojiangxia@kuaishou.com}

\author{Shuang Yang}
\affiliation{
  \institution{Kuaishou Technology}
  \city{Beijing}
  \country{China}   
}
\email{yangshuang08@kuaishou.com}

\author{Zhaojie Liu}
\affiliation{
  \institution{Kuaishou Technology}
  \city{Beijing}
  \country{China}  
}
\email{zhaotianxing@kuaishou.com}

\author{Jiawei Sheng}
\affiliation{
  \institution{Institute of Information Engineering, Chinese Academy of Sciences  \\ School of Cyber Security, University of Chinese Academy of Sciences}
  \city{Beijing}
  \country{China}
}
\email{shengjiawei@iie.ac.cn}

\author{Tingwen Liu}
\authornotemark[1]
\affiliation{
  \institution{Institute of Information Engineering, Chinese Academy of Sciences  \\ School of Cyber Security, University of Chinese Academy of Sciences}
  \city{Beijing}
  \country{China}
}
\email{liutingwen@iie.ac.cn}

\renewcommand{\shortauthors}{Ruochen Yang, et al.}

\begin{abstract}
Modern industrial recommendation systems typically separate recall and ranking into two independent stages.
Although this cascade supports corpus-level retrieval and fine-grained multi-objective scoring, it causes objective inconsistency, information loss at the candidate hand-off, and redundant user-side context computation.
Meanwhile, the generative recall and ranking scaling share a common Transformer-based modeling philosophy, where architectural consistency creates a natural opportunity for unified integration,.
However, direct sharing remains challenging since the two tasks require different information visibility and optimization methods.
Therefore, we propose \textbf{UniR$^2$}, a \textbf{Uni}fied decoder-only Transformer that unifies Generative \textbf{R}ecall and Multi-Objective \textbf{R}anking within a single heterogeneous sequence comprising user context, SID trajectory, and item features.
Within this sequence, the generated trajectory serves as a representation bridge between recall and ranking, where Dual-Query Prefix-Causal Attention provides task-specific visibility.
The two tasks share the base attention weights but retain separate optimization boundaries, with ranking-side LoRA preserving ranking adaptability without disrupting the generative backbone.
Extensive offline experiments on large-scale industrial data demonstrate the effectiveness and efficiency of UniR$^2$ for both recall and ranking.
Long-term online A/B tests on Kuaishou platform further show consistent positive gains, validating the practicality of unified model in large-scale recommendation systems.
\end{abstract}



\maketitle

\section{Introduction}

\begin{figure}[t!]
\begin{center}
\includegraphics[width=\linewidth]{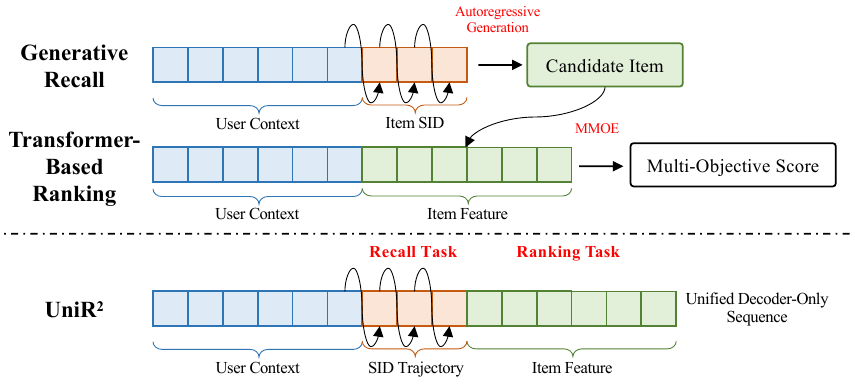}
\caption{Architecture comparison between cascaded architecture and UniR$^2$. 
Serial generation and scoring with separate models lead to misaligned objectives and redundant user context computation. 
UniR$^2$ unifies both tasks within a single heterogeneous decoder-only sequence, enabling representation reuse and bridging the cross-stage gap.}
\label{fig:motivation}
\end{center}
\end{figure}

Modern industrial recommender systems are commonly organized as a multi-stage cascade. 
Given a user request, the recall stage~\cite{youtubednn, pinsage} first retrieves thousands of candidates from a corpus containing billions of items, after which a ranking model~\cite{mmoe, sarm} performs fine-grained user-item matching under multiple business objectives. 
This division of labor has supported large-scale recommendation for years, because recall emphasizes efficient search and coverage whereas ranking emphasizes accurate per-candidate discrimination. 
However, the separation also introduces a fundamental gap between the two stages. 
Recall and ranking are usually optimized with different objectives, consume differently organized features, and produce representations in independent parameter spaces. 
As a result, information captured during candidate generation is compressed into a discrete candidate list at the stage boundary, while user context are repeatedly encoded by different models. 
These issues lead to objective inconsistency, representation loss, and redundant computation in both training and serving.

Meanwhile, both stages of this cascade are undergoing a similar architectural update toward Transformers~\cite{transformer}. 
On the recall side, generative recommendation represents each item with a hierarchical semantic ID (SID) and casts large-scale retrieval as autoregressive next-token prediction~\cite{tiger, onerec}. 
The Transformer learns a distribution over the item corpus by generating an SID trajectory conditioned on the user's behavioral context, thereby avoiding exhaustive item-by-item scoring. 
On the ranking side, recent models increasingly tokenize heterogeneous fields, behavioral sequences, and contextual features, and use multi-layer attention as a scalable feature interaction backbone~\cite{onetrans, tokenformer}. 
Although the two tasks retain different outputs, they now share several important computational primitives: tokenized inputs, user-context modeling, attention-based feature fusion, and deep Transformer stacks. 
This convergence naturally raises a question: 
\textit{\textbf{Can generative recall and multi-objective ranking be formulated into one Transformer sequence rather than two isolated models?}}

Such a unified Transformer offers several potential benefits. 
First, a common sequence space can preserve the information produced during recall and directly expose it to ranking, reducing the semantic gap at the cascade boundary. 
Besides, the two tasks can share user-side context computation, since reusing intermediate key/value caches across candidate generation and scoring eliminates the duplicated encoding of long histories and personalized profiles. 
Furthermore, scaling such a single Transformer backbone allows recall and ranking to benefit jointly, instead of independently maintaining two increasingly large models. 

Nevertheless, directly combining the two tasks is non-trivial. 
Generative recall is a structured distribution fitting problem, where each SID token must obey autoregressive causality and preserve the hierarchical semantics. 
Multi-objective ranking is instead a discriminative feature interaction problem, where all item-side features are available simultaneously, and each candidate must be evaluated under dense and sparse objectives such as click, long-view, and gift. 
Therefore, the two tasks require different information visibility and optimization methods. Simply sharing all parameters in a single sequence will lead to a severe seesaw effect between retrieval quality and ranking accuracy.

Existing attempts at recall and ranking collaboration do not fully resolve these issues. 
Some methods append a lightweight scorer after generation or transfer representations between otherwise separate modules~\cite{rankgr, oneranker}; others share a Transformer encoder or unify the input format while retaining task-specific computation paths~\cite{onepiece, unipinrec, harmonizing}. 
These approaches improve cross-stage collaboration, but the generated SID trajectory and fine-grained ranking features are still not organized within a single sequence for layer-wise interaction. 
Consequently, ranking either consumes a compressed post-generation representation or requires an additional feature-fusion network, while the user context may still be redundantly processed. 
The key challenge is therefore not simply to share an encoder or a loss function, but to construct a single sequence that provides representation coupling in the forward pass while maintaining optimization isolation between the two objectives.

To address this challenge, we propose \textbf{UniR$^2$}, a \textbf{Uni}fied decoder-only Transformer that performs Generative \textbf{R}ecall and Multi-Objective \textbf{R}anking within a single heterogeneous sequence.
UniR$^2$ organizes user context, SID trajectory, and item features in one Transformer trunk, allowing ranking to consume the recall decision process rather than only its discrete candidate output. 
A Dual-Query Prefix-Causal Attention mechanism provides task-specific visibility for causal generation and ranking feature fusion, while 
the two-stage different optimization objectives preserves ranking adaptability without destabilizing the generative backbone.
UniR$^2$ therefore achieves representation coupling and optimization isolation within one training forward pass, as well as one single service for supporting recall and ranking during online inference.

Our main contributions are summarized as follows:
\begin{itemize}
    \item We introduce a single-sequence unification perspective for generative recall and multi-objective ranking. By organizing user context, SID trajectory, and item features within one decoder-only sequence, the framework turns the recall trajectory into a representation bridge for layer-wise ranking feature interaction.
    \item We design UniR$^2$, a unified architecture equipped with DQ-PCA and LoRA-balanced optimization isolation. UniR$^2$ couples the two tasks in forward representation learning while preventing ranking gradients from corrupting the autoregressive recall backbone, and supports efficient single-service inference deployment through shared context computation.
    \item We conduct extensive experiments on large-scale industrial data. UniR$^2$ consistently improves both generative recall and multi-objective ranking over deployed baselines, and achieves significant gains in online A/B test on real-world Kuaishou services, demonstrating its effectiveness and practicality for industry.
\end{itemize}

\section{Related Work}

\subsection{Generative Recommendation}

Generative recommendation reformulates sequential recommendation as a subsequent SID prediction and generation task conditioned on users’ historical interaction sequences. 
It enables direct hierarchical search from user context to massive item candidates without scoring the entire item corpus one by one, making it naturally suitable for large-scale recall in industrial recommender systems.
This paradigm was first established by TIGER~\cite{tiger} with an item tokenizer and an encoder–decoder architecture, and was later extended by the OneRec series~\cite{onerec, oneloc, onelive, onemall}, which introduced reinforcement learning for personalized preference alignment. However, the simple autoregressive generation and beam search does not fully address the need to directly compare candidates and explore higher value ones.
PROMISE~\cite{promise} introduces a process reward model to perform fine-grained evaluation of intermediate generation paths. RankGR~\cite{rankgr} refines candidates with a lightweight scoring model. V-Star~\cite{v-star} focuses the search on critical SID branches through a structured sampling strategy. Nevertheless, these methods essentially rely on a single preference signal and do not explicitly model candidates' multi-objective scores required in the ranking stage.

\subsection{Transformer-Based Ranking}

Industrial ranking models have long relied on multi-task tower architectures~\cite{mmoe, home, ple}, achieving strong performance through manually designed feature-interaction mechanisms~\cite{dcn, din}. However, modeling user interests from different perspectives with independent networks often leads to shallow interactions among heterogeneous information, fragmented computation graphs, and limited scalability of model capacity.
In recent years, inspired by Transformer architectures~\cite{transformer}, industrial ranking models have started to represent discrete fields, contextual features, and user behavior sequences uniformly as tokens. Multi-layer attention is then used to simultaneously perform cross-field feature interaction and interest modeling, gradually endowing ranking models with an LLM-like unified backbone and scalable model capacity~\cite{wukong, longer}. OneTrans~\cite{onetrans} jointly models sequential and non-sequential features through a Transformer backbone. MixFormer~\cite{mixformer} co-scales dense feature interaction and sequence modeling within a single parameter space. SORT~\cite{sort} significantly improves the computational efficiency of ranking Transformer models through system-level optimization. TokenFormer~\cite{tokenformer} further addresses the issue of sequence representation collapse, enhancing the discriminability of multi-field feature representations while preserving sequence modeling capability.
Nevertheless, these methods still operate only at the ranking stage. Since they share the Transformer-based modeling philosophy with generative recall, there remains substantial potential to bridge recall and ranking within a unified model.

\subsection{Unified Recall and Ranking}

Collaborative modeling of recall and ranking has recently emerged as an industry-relevant research direction, evolving from objective alignment and parameter sharing toward the integration of models and computation pipelines. Early work such as RankGR~\cite{rankgr} introduces a candidate scoring module into generative recall and jointly optimizes sparse parameters, yet the two stages remain separated at the model level, which in essence is mapping recall and ranking into the same representation space.
Subsequent studies have increasingly explored unified architectures. OnePiece~\cite{onepiece} applies unified tokenized inputs, Transformer architectures, and contextual reasoning to both recall and ranking. OneRanker~\cite{oneranker} achieves coarse-to-fine collaborative optimization through key/value transfer between generation and ranking, together with distribution-consistency constraints. UniPinRec~\cite{unipinrec} unifies input formats and model training by sharing a Transformer.
However, existing methods typically rely on post-generation scoring or cross-module information transfer, and have not yet organized recall trajectories and fine-grained ranking features within the same sequence for layer-wise interaction. In contrast, while sharing the backbone network between recall and ranking objectives, UniR$^2$ integrates user context encoding, SID generation trajectories, and feature interactions into a single decoder sequence, achieving unification at both the token-sequence level and the layer-wise computation level.

\section{Preliminary}

Let $\mathcal{U}$ and $\mathcal{V}$ denote the user and item sets. 
Each user $u \in \mathcal{U}$ is associated with personalized profile and multiple history sequences of diverse behaviors.
Besides, each item $v \in \mathcal{V}$ is described by a set of raw features, containing semantic ID (SID) $s_v = \{q_1, q_2, \dots, q_L\}$ obtained through discrete quantization~\cite{tiger, onerec, onelive}, where $q_l$ is the code of the $l$-th layer. We employ Res-Kmeans~\cite{qarm} with $L=3$ layers and codebook size $8129$ in our scenario.
For every observed interaction pair $(u,v)$, we collect a multi-objective label vector $\mathbf{y}=(y_1,\dots,y_T)\!\in\!\{0,1\}^T$ ($e.g.$, click, long-view, gift) based on whether the corresponding behavior occurred.

Therefore, our goal is to accomplish two tasks on the same model foundation.
\textbf{Generative Recall} task learns an autoregressive distribution over SIDs conditioned on the user-side context:
\begin{equation}
p_\theta\bigl(s_v \, \big| \, u)=\prod_{i=1}^{L} p_\theta\bigl(q_i \, \big| \, u, q_{<i}).
\end{equation}
NTP loss is adopted during training, while beam search is performed at serving time. 
\textbf{Multi-Objective Ranking} task then learns per-objective probabilities for a given $(u,v)$ pair.
Formally, we seek a score function:
\begin{equation}
\hat{y}=f_\phi\bigl(u,v)\in[0,1]^T,
\end{equation}
whose $t$-th coordinate $\hat y_t$ estimates $P(y_t=1\mid u,v)$. 
The training of ranker is based on multi-objective BCE loss.

Traditional cascade methods instantiate the two tasks with independent parameter sets $\theta$ and $\phi$, and pass the candidate list produced by the generation task to the ranking task for multi-objective estimation. 
In this paper, we study an alternative formulation that learns a single model parameterized by $\Theta$ that can produce both the generative distribution and the ranking scores within a single stage of the pipeline. 
In this way, the candidates and their corresponding scores can be obtained in one service pass. 

\section{Methodology}\label{sec:method}

\begin{figure*}[t!]
\begin{center}
\includegraphics[width=0.95\linewidth]{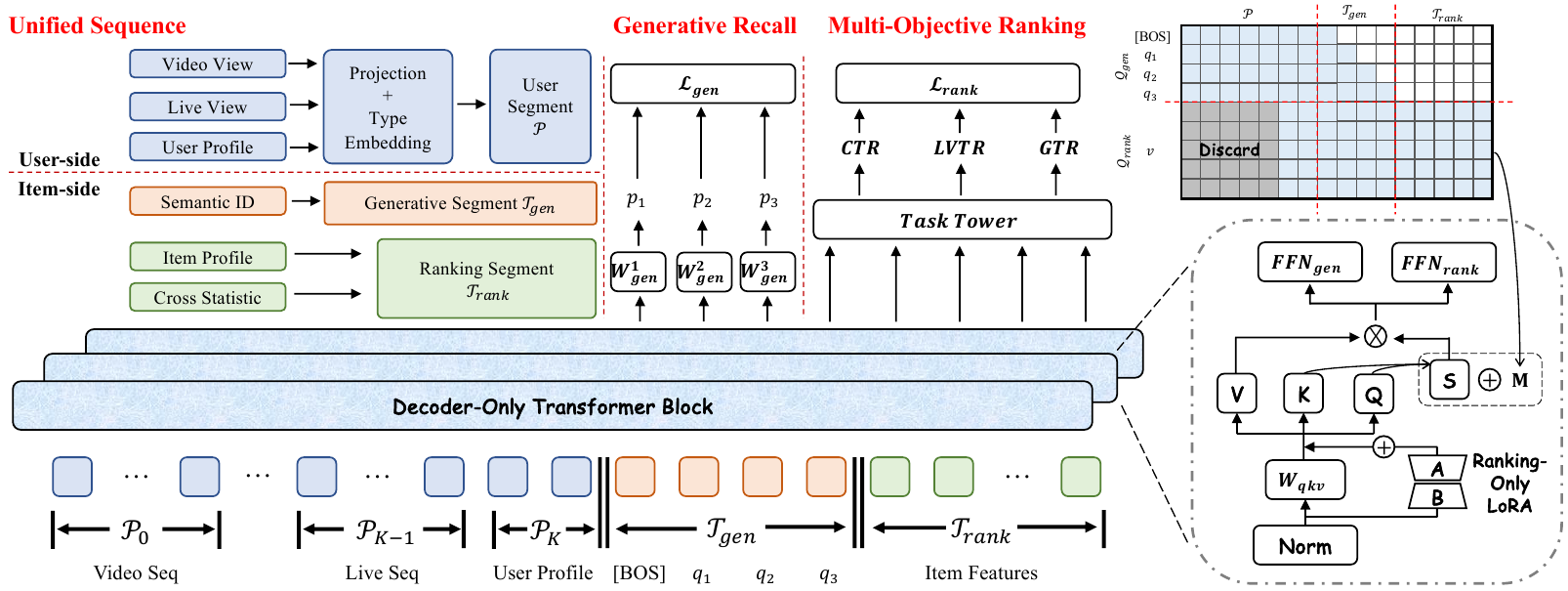}
\caption{The overall architecture of our model.}
\label{fig:model}
\end{center}
\end{figure*}

UniR$^2$ is a single decoder-only transformer that unifies generative recall and multi-objective ranking in one sequence,
thereby maximizing model reusage and reducing computational redundancy. 
As illustrated in Figure \ref{fig:model}, the input of UniR$^2$ is a sequentially arranged heterogeneous token sequence encompassing both user and item side information. 
Through a carefully designed strategy, the outputs of different segments are respectively dedicated to the generation and ranking tasks.

\subsection{Unified Sequence Composition}\label{sec:seq}

For each observable $(u,v)$ interaction pair, all relevant features are available during training. We therefore assemble these features into a single unified sequence $\mathcal{S}=[\mathcal{P}\,\|\,\mathcal{T}_{\mathrm{gen}}\,\|\,\mathcal{T}_{\mathrm{rank}}]$ that concatenates three groups of tokens as the model input.

\textbf{User segment $\mathcal{P}$.} 
We view $u$'s side information as a set of heterogeneous behavioral and profile blocks $\{\mathbf{x}_u^{(0)},\mathbf{x}_u^{(1)},\dots,\mathbf{x}_u^{(K-1)},\mathbf{x}_u^{(K)}\}$, where $\mathbf{x}_u^{(K)}$ encodes $u$'s static profile and $\mathbf{x}_u^{(b)}$ ($0\le\!b\!\le K-1$) is a variable-length interaction history sequence under behavior 
$b \in \{\text{Click}, \text{Long-View}, \text{Gift}, \dots\}$.
Each block is projected to $d$ and shifted by its own type embedding drawn from a table $E_{\mathcal{P}}\!\in\!\mathbb{R}^{(K+1)\times d}$:
\begin{equation}
\mathcal{P}_k=\mathrm{Proj}_k(\mathbf{x}_u^{(k)})+E_{\mathcal{P}}[k],\quad
\mathcal{P}=[\mathcal{P}_0;\mathcal{P}_1;\dots;\mathcal{P}_{K-1};\mathcal{P}_K].
\end{equation}
Since $\mathcal{P}$ represents stable user side information and remains unchanged across different items, it serves as a shared memory read by each downstream target token, playing the role of an encoder key/value cache.

\textbf{Generative segment $\mathcal{T}_{\mathrm{gen}}$.} 
Given the target item $v$ with hierarchical SID $s_v = (q_1, q_2, q_3)$, this segment is the teacher-forced input of the generative recall task:
\begin{equation}
    \mathcal{T}_{\mathrm{gen}} = [\textsc{[bos]}, e(q_1), e(q_2), e(q_3)],
\end{equation}
where $e(\cdot)$ is the codebook embedding lookup and $\textsc{[bos]}$ is a learned special token. 
Position $i$ of $\mathcal{T}_{\mathrm{gen}}$ is trained to predict the next-level code $q_i$ in the standard NTP pattern.

\textbf{Ranking segment $\mathcal{T}_{\mathrm{rank}}$.} 
The SID alone is insufficient to fully model item characteristic or feature cross between $(u,v)$ pair. Therefore, in addition to the SID of $v$, we append an ordered set of item feature tokens:
\begin{equation}
    \mathcal{T}_{\mathrm{rank}}=[e_1(v),\,e_2(v),\,\dots,\,e_M(v)],
\end{equation}
which summarizes $v$ from complementary views, \textit{e.g.} item profile, content embedding, $(u,v)$ cross statistics and upstream priors.
Each $e_m(v)$ is produced by an independent lightweight projection over its raw feature block and shifted by a position type embedding. 
These tokens act as learnable ranking queries whose contextualized representations
are used as sufficient feature fusion results to be transmitted to the multi-objective ranking tower.

Therefore, the complete input single sequence is:
\begin{equation}
\mathcal{S}=\bigl[\underbrace{\mathcal{P}_0,\dots,\mathcal{P}_{K-1},\mathcal{P}_K}_{\text{User segment}}\;\bigl\|\;\underbrace{\textsc{[bos]},e(q_1),e(q_2),e(q_3)}_{\text{Generative segment}}\;\bigl\|\;\underbrace{e_1(v),\dots,e_M(v)}_{\text{Ranking segment}}\bigr].
\end{equation}
This unified sequence composition design allows the recall and ranking targets to be fed into the same decoder-only Transformer block, where they sit within the same stack and share both model parameters and the contextual memory of invariant content.

\subsection{Dual-Query Prefix-Causal Attention}\label{sec:mask}

A naïve practice for Transformer-based generative recall~\cite{onerec-v2, nezha} or ranking~\cite{onetrans, loopctr} is to apply causal self-attention over the entire unified sequence. 
While conceptually straightforward, this design is computationally wasteful. 
First, the user segment is typically much longer than the two target segments, making it highly redundant to force the full sequence to perform attention at every layer. 
Second, the two target segments play  roles, since $\mathcal{T}_{\mathrm{gen}}$ follows an autoregressive tasks and therefore cannot access future positions, whereas  $\mathcal{T}_{\mathrm{rank}}$ consists of item-side information added synchronously and can thus be mutually visible.
To address this, UniR$^2$ introduces a Dual-Query Prefix-Causal Attention mechanism (DQ-PCA), which keeps the two objectives separated along the sequence axis while sharing all base attention weights.

\textbf{Dual queries with limited visibility.}
At each layer, for the two distinct target segments, the block consumes the unified sequence using two independent queries with different visibility patterns:
\begin{itemize}
    \item The generative query $Q^{\mathrm{gen}}$ carries the SID tokens $\mathcal{T}_{\mathrm{gen}}$. Consistent with standard generative recall, predicting possible SID only depends on user-side context. Therefore $Q^{\mathrm{gen}}$ attend over the full user segment and over its own preceding positions, implementing a prefix-causal pattern:
    \begin{equation}
        Q^{\mathrm{gen}} = [\mathcal{T}_{\mathrm{gen}}], \quad K^{\mathrm{gen}} = V^{\mathrm{gen}} = [\mathcal{P}, \mathcal{T}_{\mathrm{gen}}].
    \end{equation}
    Formally, indexing SID query positions by $i\!\in\![1,L]$ and key positions by $j\!\in\![1,|\mathcal{P}|+L]$, the attention mask reads:
    \begin{equation}
        \mathbf{M}^{\mathrm{gen}}_{i,j}=
        \begin{cases}
        0, & j\le|\mathcal{P}|\ \ \text{or}\ \ |\mathcal{P}|<j\le|\mathcal{P}|+i,\\[2pt]
        -\infty, & |\mathcal{P}|+i<j\le|\mathcal{P}|+L.
        \end{cases}
    \end{equation}
    \item The ranking query $Q^{\mathrm{rank}}$ carries the item-side feature tokens $\mathcal{T}_{\mathrm{rank}}$. Since every element of the unified sequence is in principle mutually informative for scoring, bidirectional self-attention is the most expressive choice for feature fusion.
    However, considering that the long user history has already been amortized by the generative stage, and the sequential retrieval signals such as GSU~\cite{sim} results are already encoded as tokens inside $\mathcal{T}_{\mathrm{rank}}$ itself. Therefore $Q^{\mathrm{rank}}$ attend the partial sequence within the user segment is discarded while only the user profile is retained:
    \begin{equation}
        Q^{\mathrm{rank}} = [\mathcal{T}_{\mathrm{rank}}], \quad K^{\mathrm{rank}} = V^{\mathrm{rank}} = [\mathcal{P}_{K}, \mathcal{T}_{\mathrm{gen}}, \mathcal{T}_{\mathrm{rank}}].
    \end{equation}
    The additive mask is uniformly zero within constrained scope:
    \begin{equation}
        \mathbf{M}^{\mathrm{rank}}_{i,j}=0,\quad \forall\,i\!\in\![1,M],\ j\!\in\![1,|\mathcal{P}_{K}|+L+M].
    \end{equation}
\end{itemize}

It can be observed that the personalized identity information in the prefix user segment is read by both queries, which avoids redundant computation within a single forward pass. 
In addition, the ranking segment attends to the complete generative segment, so the ranking output is always produced from a representation that has already incorporated the recall trajectory corresponding to the target item. 
These make UniR$^2$ a genuinely unified model, rather than two co-located submodels sharing the same encoder, and it delivers representation coupling by construction.

\textbf{Weights sharing.}
Similar to the normal decoder-only architecture, each block of UniR$^2$ consists of an attention mechanism and a Feed-Forward Network.
The former is based on multi-head attention and uses DQ-PCA to construct corresponding inputs for different targets:
\begin{equation}
    O = \text{MultiHead}(W_qQ, W_kK, W_vV)W_o,
\end{equation}
The base projections $W_q,W_k,W_v,W_o\in \mathbb{R}^{d\times d}$ are shared between the two branches and across all memory partitions.

The outputs corresponding to different queries are then passed through task-specific feed-forward blocks $\text{FFN}_{gen}$ and $\text{FFN}_{rank}$, thereby establishing a task-level barrier between the two objectives at each layer~\cite{sensenova}.
Stacking $N$ layers of such blocks yields $\mathbf{h}^{\mathrm{gen}}\!\in\!\mathbb{R}^{L\times d}$ for the recall head and $\mathbf{h}^{\mathrm{rank}}\!\in\!\mathbb{R}^{M\times d}$ for the ranking head.
In this way, the architecture fully preserves a shared backbone and the invariance of a single forward pass, which together serve as the foundation for unification.

\subsection{Generative Recall}

The generative branch aims at turning generative hidden states $\mathbf{h}^{\mathrm{gen}}$ into a distribution over the hierarchical SID codebook. To let each SID position specialize to its own codebook level while still sharing the transformer trunk, we attach $L$ independent output projections $W_{gen}^{(1)},\dots,W_{gen}^{(L)}\!\in\!\mathbb{R}^{V\times d}$ on top of $\mathbf{h}^{\mathrm{gen}}$ and predict the code of the next level $i$ from the preceding hidden state:
\begin{equation}
    p_{\Theta}\bigl(q_i\,\big|\,u,q_{<i}\bigr) =\mathrm{softmax}\bigl(W_{gen}^{(i)}\,\mathbf{h}^{\mathrm{gen}}_{i-1}\bigr).
    \label{eq:sid_head}
\end{equation}
The generation task is optimized based on a weighted cross-entropy:
\begin{equation}
\label{eq:grm_loss}
    \mathcal{L}_{\mathrm{gen}}(\Theta)
    =\sum_{i=1}^{L}\alpha_i\bigl[-\log p_{\Theta}\bigl(q_i\,\big|\,u,q_{<i}\bigr)],
\end{equation}
with weights $\alpha_1\!\ge\!\cdots\!\ge\!\alpha_L\!>\!0$ since the code placed ahead can determine the attributes of a generated item from a higher dimension.
Because $\mathcal{L}_{\mathrm{gen}}$ only reads $\mathbf{h}^{\mathrm{gen}}$, its gradients flow exclusively through $Q^{\mathrm{gen}}$'s attention paths and never reach the item-side components.

\subsection{Multi-Objective Ranking}\label{sec:rank}

The ranking branch uses the contextualized ranking hidden states $\mathbf{h}^{\mathrm{rank}}$ to estimate multiple business objectives for the candidate items. 
Since the ranking model in recommendation expects multi-source complementary signals as input, $\mathbf{h}^{\mathrm{rank}}$ only preserves target-aware deep interaction representations. 
Therefore, we additionally incorporate the original user and author features to prevent shallow signals from being smoothed out. 
Moreover, we also concatenate the output at the last SID position $\mathbf{h}^{\mathrm{rank}}_{L}$. 
Although the ground-truth SID at this position is redundant for autoregressive computation in the generative task, this hidden state observes the entire generation trajectory and thus provides target semantics from a perspective different from the discriminative representation. 
Therefore, the final input to the ranker consists of:
\begin{equation}
    z = \text{Concat}(u, v, \mathbf{h}^{\mathrm{gen}}_{L}, \mathbf{h}^{\mathrm{rank}}).
\end{equation}
The vector $z$ is fed into an MMoE~\cite{mmoe} model, where each task-specific tower outputs $\hat{y}_t$ for objective $t$. The entire ranker is trained with the BCE loss:
\begin{equation}
\label{eq:rank_loss}
    \mathcal{L}_{\mathrm{rank}}(\Theta)
    = \sum_{t=1}^{T}\omega_t\cdot
    \mathrm{BCE}\bigl(\hat{y}_t,y_t\bigr),
\end{equation}
where $\omega_t$ is the objective weight.

\textbf{Gradient entanglement.}
In this way, UniR$^2$ completes the prediction of both generative recall and ranking within a single forward pass over one unified sequence. 
However, if we simply define the overall training objective as the sum of the two tasks:
\begin{equation}
\label{eq:loss}
    \mathcal{L}(\Theta) = \mathcal{L}_{\mathrm{gen}}(\Theta)+\mathcal{L}_{\mathrm{rank}}(\Theta),
\end{equation}
the gradients from both generative recall and ranking will be propagated into shared parameters, which can introduce cause optimization interference.

A straightforward solution is to apply stop-gradient before the ranking head: $\tilde{z} = \mathrm{sg}(z)$.
This blocks $\mathcal{L}_{\mathrm{rank}}$ from back-propagating into the Transformer trunk and protects the generative path. However, it also freezes ranking-side feature fusion: the head can only learn on top of a representation mainly optimized for SID prediction. 
This often amplifies the seesaw effect, because stronger objectives dominate the head while weaker or delayed objectives cannot adapt the attention-level interactions. 
UniR$^2$ therefore keeps the stop-gradient boundary for safety, but adds a learnable low-rank path activated only for the ranking view.

\textbf{LoRA-balanced ranking adaptation.}
We inject LoRA~\cite{lora} into the attention projections used by the ranking branch. The base projections $W_q,W_k,W_v$ remain shared with the generative branch, but their outputs are detached when they are consumed by the ranking query. 
Formally, ranking-specific adaptation is carried by low-rank residual matrices:
\begin{equation}
    \Delta W^{\mathrm{rank}}_X = B^{\mathrm{rank}}_XA^{\mathrm{rank}}_X,
    \quad X^{\mathrm{rank}}=\mathrm{sg}(WX)+\Delta W^{\mathrm{rank}}_XX,
\end{equation}
where $X\in \{Q,K,V\}$ is the inputs of the ranking branch, and $B^{\mathrm{rank}}_X\in\mathbb{R}^{d\times r},A^{\mathrm{rank}}_X\in\mathbb{R}^{r\times d}$ are low-rank parameters. These modules take effect only in ranking-view attention in each layer. The generative query still uses the original base projections and never activates the ranking LoRA branch. Thus, LoRA is placed at the Q/K/V projections that determine ranking-side feature fusion, giving the ranker enough adaptive capacity while isolating the base generative backbone from ranking gradients.

In general, trainable parameters are divided into generative, ranking, and shared sparse embedding parts:
\begin{equation}
    \Theta_{\mathrm{gen}}\cap\Theta_{\mathrm{rank}}=\varnothing,
    \quad   \Theta=\Theta_{\mathrm{gen}}\cup\Theta_{\mathrm{rank}}\cup\Theta_{\mathrm{emb}}.
\end{equation}
$\Theta_{\mathrm{gen}}$ contains the base Transformer projections, generative FFN, and recall heads; $\Theta_{\mathrm{rank}}$ contains ranking LoRA matrices, ranking-side FFN parameters, and multi-task towers; $\Theta_{\mathrm{emb}}$ contains sparse embeddings shared by both tasks~\cite{grank}. 
UniR$^2$ therefore preserves representation coupling in the forward pass, while tasks separated in the optimization and modules.

\section{Deployment}

\subsection{Training Strategy}

\textbf{Data sample.}
Mainstream generative recall methods essentially perform maximum-likelihood fitting over the online exposure distribution. Their training sets usually contain only exposed or positive feedback samples, without explicit discriminative negative samples. 
However, the ranker inside UniR$^2$ is jointly trained with the generative branch in the same forward pass. 
If we directly inherit the sampling strategy of generative recall, each XTR objective would lack negative contrast, leading to degraded scoring ability.

Therefore, we adopt full sample stream with target-adaptive masks strategy. 
All $(u,v)$ pair samples entering the model are visible to the ranker, while each discriminative objective uses an independent sample mask and contributes loss only on the subset semantically relevant to that objective. 
This provides sufficient contrast for all objectives within the same batch.
Above ranker stream, the generative branch further performs downsampling, which keeps all clicked samples and randomly exposed samples as positive samples for the GRM loss,
so as to preserve the recall-side semantics of fitting the online exposure distribution. 
These two sampling strategies coexist within the same forward pass, and control whether to participate in the loss through their respective masks, thereby balancing the discriminative contrast required for ranking and the distribution fitting target required for recall.

\textbf{Two-Stage Training.}
We adopt a two-stage joint training strategy. In the first stage, only the GRM loss in Eq.\ref{eq:grm_loss} is enabled to obtain a backbone with stable distribution fitting. In the second stage, training is warm-started from this checkpoint, and the ranker loss is appended on the same sequence as Eq.\ref{eq:loss} for joint training.

This design is mainly motivated by the fact that the discriminative input of the ranker depends on the SID representations produced by the generative branch. 
If the ranker is introduced at the early stage when the SID semantics are still unstable and drifting, it can be misled by substantial noise and form incorrect decision boundaries, resulting in negative transfer. 
Therefore, the two-stage design decouples distribution fitting and discriminative alignment in the optimization space, imposing a curriculum prior on multi-task learning.

\subsection{Inference Pipeline}

\begin{figure}[t!]
\begin{center}
\includegraphics[width=\linewidth]{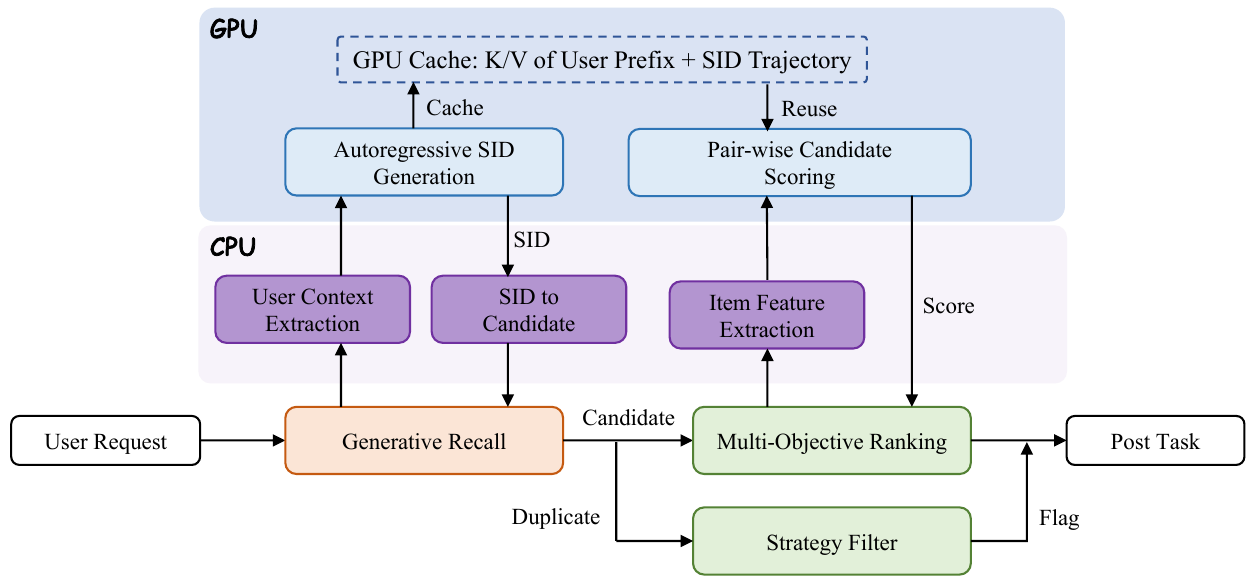}
\caption{The inference pipeline of UniR$^2$. GPU caches the user context and SID-trajectory KV states computed during recall for direct reuse in ranking, while strategy filtering and ranking are executed in parallel.}
\label{fig:inference}
\end{center}
\end{figure}

\begin{table*}[t]
  \caption{The overall recall performance comparison of different models on live-streaming offline dataset. The best results are \textbf{boldfaced} and the second-best results are \underline{underlined}, with the relative improvements denoted as \textit{Imprv.}$\uparrow$.}
  \label{tab:offline_recall}
  \begin{tabular}{lcccccccc}
    \toprule
    \multirow{2.5}{*}{\textbf{Models}} & \multicolumn{4}{c}{\textbf{Show}} & \multicolumn{4}{c}{\textbf{Click}} \\
    \cmidrule(r){2-5} \cmidrule(r){6-9} & \textbf{HR@64} & \textbf{MRR@64} & \textbf{HR@128} & \textbf{MRR@128} & \textbf{HR@64} & \textbf{MRR@64} & \textbf{HR@128} & \textbf{MRR@128} \\ 

    \midrule
    \midrule
    KuaiFormer & 0.4176 & 0.1328 & 0.4907 & 0.1337 & 0.4479 & 0.1499 & 0.5164 & 0.1507  \\
    GNN & 0.4552 & 0.1532 & 0.5269 & 0.1540 & 0.5463 & 0.1941 & 0.6197 & 0.1949  \\

    \midrule
    OneRec & 0.6844 & 0.3224 & 0.7261 & 0.3230 & 0.7494 & 0.3987 & 0.7789 & 0.3992 \\
    OneLive & 0.7177 & 0.3254 & 0.7742 & 0.3260 & 0.7799 & 0.4046 & 0.8284 & 0.4051  \\
    NEZHA & 0.7244 & 0.3221 & 0.7794 & 0.3228 & 0.7848 & 0.4062 & 0.8312 & 0.4067  \\
    PROMISE & \underline{0.7329} & \underline{0.3391} & \underline{0.7874} & \underline{0.3396} & \underline{0.7932} & \underline{0.4286} & \underline{0.8416} & \underline{0.4289}  \\

    \midrule
    UniR$^2$ & \textbf{0.7662} & \textbf{0.3506} & \textbf{0.8114} & \textbf{0.3563} & \textbf{0.8309} & \textbf{0.4441} & \textbf{0.8741} & \textbf{0.4516}  \\
    \textit{Imprv.}$\uparrow$ & +4.54\% & +3.39\% & +3.05\% & +4.92\% & +4.75\% & +3.62\% & +3.86\% & +5.29\% \\
    
    \bottomrule
  \end{tabular}
\end{table*}

\begin{table}[t]
  \caption{The overall rank performance comparison of UniR$^2$ with baseline on live-streaming offline dataset.}
  \label{tab:offline_rank}
  \begin{tabular}{lcccccc}
    \toprule
    \multirow{2.5}{*}{\textbf{Models}} & \multicolumn{2}{c}{\textbf{CTR}} & \multicolumn{2}{c}{\textbf{LVTR}} & \multicolumn{2}{c}{\textbf{GTR}} \\
    \cmidrule(r){2-3} \cmidrule(r){4-5} \cmidrule(r){6-7} & \textbf{AUC} & \textbf{UAUC} & \textbf{AUC} & \textbf{UAUC} & \textbf{AUC} & \textbf{UAUC} \\ 

    \midrule
    Base & 0.8450 & 0.6635 & 0.8994 & 0.7591 & 0.9541 & 0.6589 \\
    UniR$^2$ & 0.8513 & 0.6731 & 0.9096 & 0.7623 & 0.9556 & 0.6601 \\
    \textit{Imprv.}$\uparrow$ & +0.75\% & +1.45\% & +1.13\% & +0.42\% & +0.16\% & +0.18\% \\

    \bottomrule
  \end{tabular}
\end{table}

In a conventional industrial recommender system, a user request is first processed by a recall model to retrieve hundreds of candidates from the entire item corpus.
After candidate-level features are extracted, an independent ranking model evaluates each user-item pair and produces multi-objective scores.
Between the recall and ranking processes, the generated candidates should undergo further processing through commercial strategies such as qualification review, deduplication, and content filtering before being passed on to the downstream stages.
This pipeline alternates between CPU and GPU execution.
Feature extraction and service orchestration are mainly performed on CPUs, whereas the forward passes of the recall and ranking models are executed on GPUs.
Because recall and ranking are deployed as independent services, the ranking stage needs to reconstruct the user context and perform another complete model forward pass, introducing redundant user-side computation, cross-service data transfer, and additional scheduling overhead.

UniR$^2$ replaces the cascaded recall and ranking models with a single serving instance.
Given a user request, the service first extracts the user profile and historical behavior features and then performs autoregressive SID generation.
During recall, the key and value representations of the user prefix, as well as the KV states corresponding to SID generation trajectory of each candidate $v$, are stored in the GPU KV cache:
\begin{equation}
    \mathcal{C}_{u,v}
    =
    \left\{
    K_{\ell}(\mathcal{P}),
    V_{\ell}(\mathcal{P})
    \right\}_{\ell=1}^{N}
    \cup
    \left\{
    K_{\ell}(\mathcal{T}_{\mathrm{gen}}^{v}),
    V_{\ell}(\mathcal{T}_{\mathrm{gen}}^{v})
    \right\}_{\ell=1}^{N},
\end{equation}
where $N$ is the number of Transformer layers.

After recall, the CPUs extract the item-side attributes of the retrieved candidates.
These attributes are projected into ranking tokens and appended to the cached unified sequence.
The ranking query can therefore directly attend to the cached user context and the corresponding SID trajectory without recomputing either segment.
The candidate ranking stage only needs to compute the newly appended item-feature tokens and ranking-specific LoRA paths.
Moreover, due to the full view, there is no need for autoregression and only a single sequence calculation is required.

A direct implementation could sequentially execute recall, strategy filtering, and ranking.
However, in our production scenario, the external strategy service takes longer than the UniR$^2$ ranking forward pass, which makes sequential execution offset the latency benefit of model unification.
To avoid this redundant time, we duplicate the recalled candidate set and dispatch it to two parallel branches.
The strategy branch transfers the candidate attributes to the filtering service and produces a binary filter flag for each candidate.
Meanwhile, the ranking branch extracts item-side features
and computes multi-objective scores for all candidates in a batched GPU forward pass.
Once ranking is completed, it waits only for the filter flags and removes candidates that violate the strategy constraints, hiding ranking computation behind the strategy latency.
This single-service deployment simultaneously eliminates repeated user-context computation, preserves the recall trajectory for ranking, and overlaps model inference with external strategy execution.
The pipeline is illustrated in Figure \ref{fig:inference}.

\section{Experiments}

In this section, we conduct extensive experiments offline and online to answer the following questions:
\begin{itemize}
    \item \textbf{RQ1:} How does our unified model perform compared to the recall and ranking baselines?
    \item \textbf{RQ2:} How does the proposed key structural components contribute to the overall model performance?
    \item \textbf{RQ3:} How does our unified model perform in  online services?
\end{itemize}

\subsection{Experimental Settings}

\subsubsection{\textbf{Dataset.}}

We conduct our evaluation based on large-scale industrial live-streaming platform logs from Kuaishou App.
This dataset includes 400 million users and 3 million authors, as well as billions of diverse interactions among them.
For generative recall task, we mainly use the exposure (show) and click records for test.
For multi-objective ranking task, we mainly report the core objectives such as click, long-view and gift, with the corresponding through rate denoted as CTR, LVTR and GTR.

\subsubsection{\textbf{Evaluation Metrics.}}

The evaluation considers both recall and ranking performance, therefore we adopt Accuracy (ACC), Hit Rate (HR) and Mean Reciprocal Rank (MRR) for recall task and AUC, UAUC for ranking task, which are all widely used offline evaluation metrics in prior works~\cite{onerec, onelive, sim, onetrans}.

\subsubsection{\textbf{Baselines.}}

For the recall task, we compare UniR$^2$ with two groups of baselines: (i) traditional recall methods, including the long-sequence model KuaiFormer~\cite{kuaiformer} and graph-structured model GNN~\cite{lightgcn}; (ii) generative methods, including OneRec~\cite{onerec} and OneLive~\cite{onelive}, as well as the efficient decoding framework NEZHA~\cite{nezha} and the process reward model PROMISE~\cite{promise}. The latter two baselines and UniR$^2$ are all modified based on OneLive.
For the ranking task, our baseline is an MMoE-style multi-objective ranking model HoME~\cite{home} with conventional feature fusion, which has been deployed online and serves as the basic of our platform. We directly compare our unified model with the online production models.

\begin{table*}[t]
  \caption{The overall performance comparison of variants with different components ablation.}
  \label{tab:ablation}
  \begin{tabular}{llcccccccc}
    \toprule
    \multirow{2.5}{*}{\textbf{Branch}} & \multirow{2.5}{*}{\textbf{Variants}} & \multicolumn{3}{c}{\textbf{Recall}} & \multicolumn{3}{c}{\textbf{Ranking}} & \multicolumn{2}{c}{\textbf{Efficiency}} \\
    \cmidrule(r){3-5} \cmidrule(r){6-8} \cmidrule(r){9-10} & & $\mathcal{L}_{gen}\downarrow$ & HR@1$\uparrow$ & ACC@all$\uparrow$ & $\mathcal{L}_{rank}\downarrow$ & AUC@CTR$\uparrow$ & AUC@GTR$\uparrow$ & Params & FLOPs \\

    \midrule
    & UniR$^2$ & 6.8260 & 0.1826 & 0.6249 & 5.9528 & 0.8450 & 0.9541 & 75M & 9.37G \\

    \midrule
    \multirow{2}{*}{\textbf{Recall}} & \textit{w}. lazy decoder-only & +0.92\% & -1.92\% & -0.46\% & - & - & - & 74M & 7.42G \\
    & \textit{w}. causal self-attention & -0.89\% & +2.85\% & +0.46\% & - & - & - & 75M & 45.09G \\

    \midrule
    \multirow{3}{*}{\textbf{Ranking}} & \textit{w}/\textit{o}. DQ-PCA \& sg & +30.10\% & -44.39\% & -10.57\% & -3.83\% & +3.97\% & +5.82\% & 72M & 7.43G \\
    & \textit{w}/\textit{o}. history discard & +0.01\% & +0.00\% & -0.01\% & -0.34\% & +2.38\% & +2.87\% & 75M & 15.39G \\
    & \textit{w}/\textit{o}. LoRA & -0.02\% & +0.01\% & +0.04\% & +7.05\% & -13.79\% & -31.29\% & 75M & 9.27G \\

    \bottomrule
  \end{tabular}
\end{table*}

\subsection{Overall Experiments (RQ1)}

\subsubsection{\textbf{Recall Performance}}

As the results shown in Table \ref{tab:offline_recall}, UniR$^2$ consistently outperforms both conventional and generative baselines in terms of generative recall, demonstrating that the unified sequence formulation does not sacrifice the model’s core capability of generating SIDs.
This improvement can be attributed to two key factors. 
First, our carefully designed attention mechanism enables SIDs in the recall branch to effectively perceive the rich user's interest representations encoded from historical interactions through prefix-cross attention. 
Second, the dual branches design decouples gradients for model parameters, thereby mitigating the conflicting influence of the ranking objective on generative capability. 
Meanwhile, the retained shared embeddings still allow the shallow signals to benefit from fine-grained optimization through positive–negative sample contrast, which may further explain the substantial improvement in generative recall performance.

\subsubsection{\textbf{Rank Performance}}

We compare UniR$^2$ with the online ranking model in Table \ref{tab:offline_rank}. As can be observed, UniR$^2$ improves XTR prediction performance on both sparse and dense targets, demonstrating that the shared decoder-only sequence architecture can serve as an effective backbone for feature fusion.
The ranking module does not merely passively reuse the generative backbone. Instead, it preserves task-specific feature interactions for ranking through LoRA-based adaptation.
Besides, the unified design benefits from the generated SID trajectory as an additional target-aware semantic view, complementing the original ranking features while preserving the discriminative contrast required by ranking.

Overall, the offline results verify the effectiveness of UniR$^2$ on both generative recall and multi-objective ranking.
It achieves stronger candidate retrieval quality while effectively improves XTR prediction performance.
This indicates that a shared decoder-only sequence architecture can integrate distribution fitting of generation and 
contrast drive of ranking within a single model.

\subsection{In-depth Analysis (RQ2)}

\subsubsection{Components Ablation}

We conducted extensive experiments on the core components of UniR$^2$ to quantify the contributions of these designs. The results are presented in Table \ref{tab:ablation}. 
We summarize the observations as follows:
\begin{itemize}
    \item For the generative recall branch, since the lazy decoder-only architecture in the baseline~\cite{onerec-v2, onelive} decouples user history encoding from SID generation, which leads to fragmented learning and representation sharing.
    Applying full-sequence causal self-attention brings only limited improvement while introducing substantially higher computational cost.
    In contrast, UniR$^2$ achieves a more favorable effectiveness and efficiency trade-off through prefix-cross attention, which provides sufficient access to historical interests without densely encoding the entire user history sequence at every layer.
    \item For the ranking branch, removing DQ-PCA and corresponding gradient truncation causes the model to strongly tilt toward ranking after long joint training, where ranking metrics improve but generative performance drops significantly. 
    This phenomenon comes from the optimization asymmetry between the two tasks.
    Ranking is supervised by multiple dense discriminative objectives and can provide more frequent gradients, while SID generation is a structured autoregressive task that requires preserving token-level semantic consistency. 
    Without dual-query visibility control and stop-gradient isolation, ranking gradients can directly reshape the shared attention representations, making the backbone increasingly optimized for discriminative scoring rather than stable SID generation.
    \item Discarding the long history sequence for the ranking query greatly reduces FLOPs while causing little degradation in ranking performance. 
    This suggests that the ranking branch does not need to repeatedly attend to the full user history. 
    The generated SID trajectory has already condensed the user-item matching process from the recall branch, and item-side sequential features such as GSU signals further summarize similar interests. 
    Therefore, attending to the SID trajectory together with compact ranking features is sufficient for feature fusion.
    \item  The LoRA is designed to provide ranking-specific adaptation capacity under optimization isolation. 
    Gradient truncation protects the generative backbone but freeze the attention-level feature interactions available to the ranker. 
    By adding low-rank residual paths only to the ranking-view attention projections, UniR$^2$ allows the ranker to learn task-specific feature fusion without modifying the base generative parameters and introducing only a small amount of parameters and computational load.
    Its impact on sparse targets such as gift is particularly significant.
\end{itemize}

\subsubsection{Attention Analysis}

\begin{figure}[t!]
    \centering
    \subfloat[Attention Weights of Recall]{
        \includegraphics[width=8.7cm]{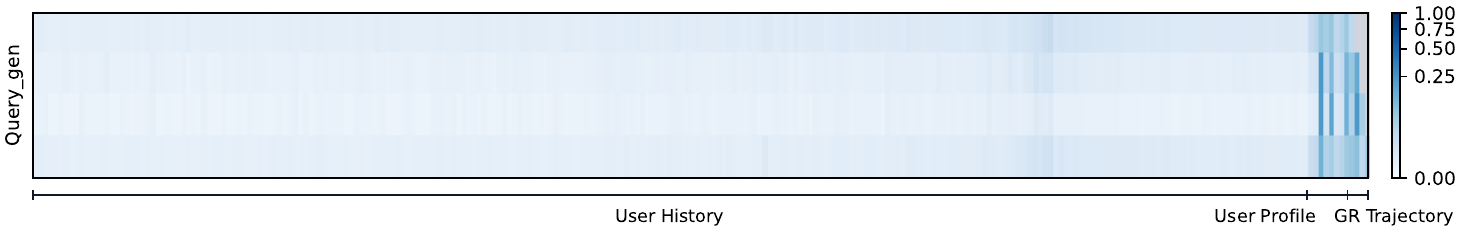}
    }
    \quad
    \subfloat[Attention Weights of Ranking \\ \quad \textit{w}/\textit{o}. LoRA]{
        \includegraphics[width=4.2cm]{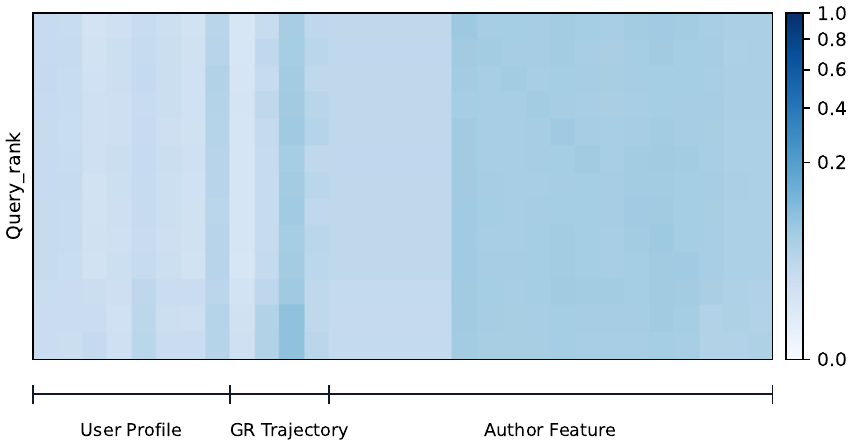}
    }
    \subfloat[Attention Weights of Ranking \\ \quad \textit{w}. LoRA]{
        \includegraphics[width=4.2cm]{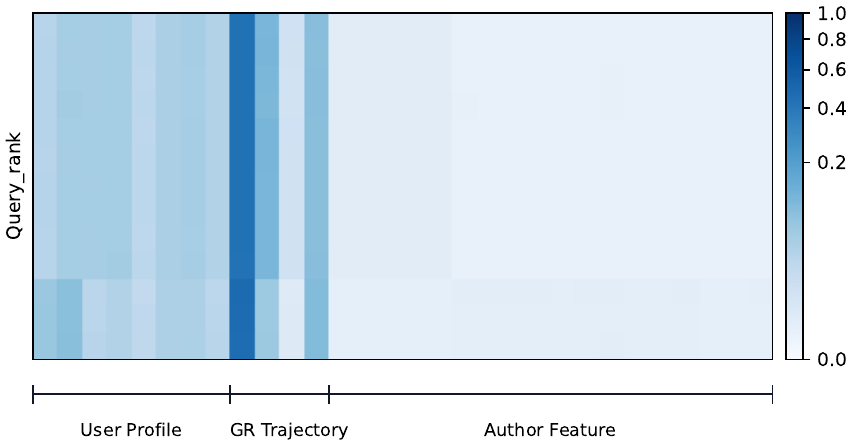}
    }
    \caption{Visualization of attention weights of recall and ranking at an intermediate layer.}
    \vspace{-0.3cm}
    \label{fig:weight}
\end{figure}

We visualize the attention weights corresponding to the recall and ranking queries in the unified sequence in Figure \ref{fig:weight}. 
For generative recall, the SID query attends broadly to the behavioral prefix. 
This pattern is consistent with the hierarchical generation process, while the historical behaviors provide fine-grained evidence of dynamic interests, the profile features offer stable personalized priors, and previously generated SID tokens constrain the semantics of the next code.

The ranking-side comparison further reveals the role of LoRA-based adaptation. 
Without LoRA, attention inherited from the recall-oriented base projections is relatively diffuse and places substantial mass on author features. 
After enabling ranking-specific LoRA, the distribution becomes much sharper around the user profile and generation trajectory.
The ranker therefore learns to emphasize target-aware user-item matching signals that have been distilled by the recall process, rather than merely reusing a generic representation learned for SID prediction. 
This verifies that the generation trajectory serves as an effective representation bridge and that LoRA restores ranking-specific feature fusion without perturbing the generative backbone.

\subsubsection{Scaling Unified Architecture}

\begin{figure}[t!]
    \centering
    \subfloat[Scaling Depth]{
        \includegraphics[width=4cm]{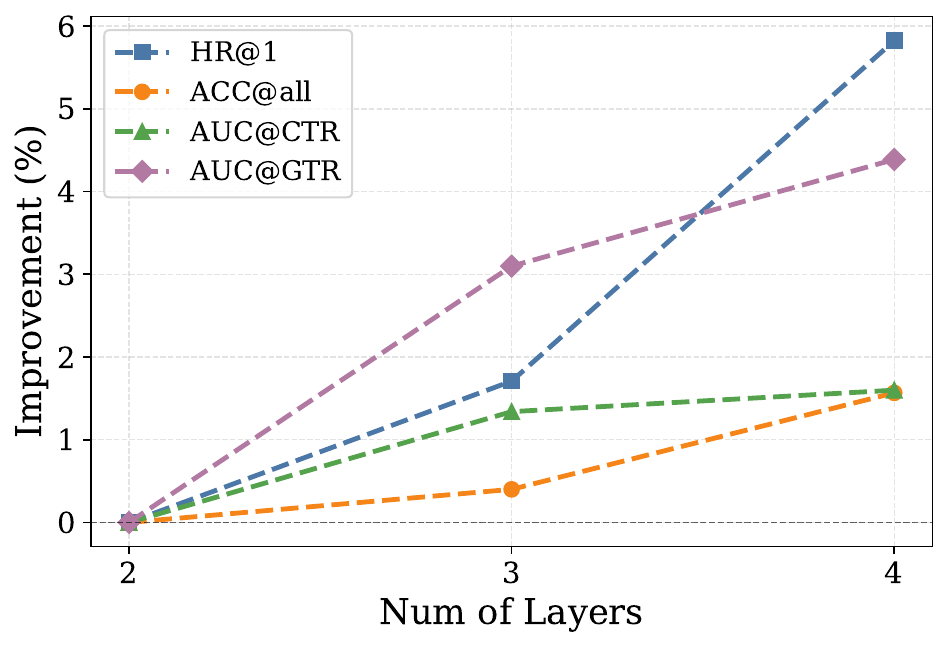}
    }
    \subfloat[Scaling Width]{
        \includegraphics[width=4cm]{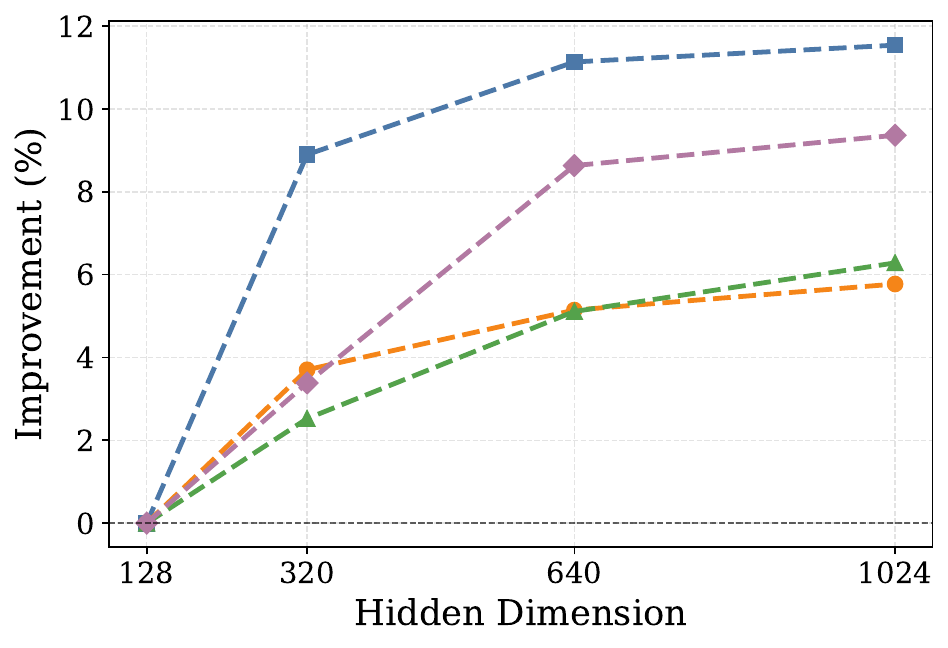}
    }
    \caption{Performance improvements obtained by scaling the depth and width of the unified architecture.}
    \label{fig:scaling}
\end{figure}

A key intuition behind the unified architecture is that both recall and ranking objectives can benefit from scaling the shared backbone. 
We therefore conduct depth and width scaling experiments, with the results shown in Figure \ref{fig:scaling}. 
As can be observed, scaling from both brings substantial improvements in generative and ranking capabilities, which verifies the strong scalability of UniR$^2$.
Notably, a larger unified backbone enhances both distribution fitting and discriminative prediction without inducing an imbalanced trade-off between them. This suggests that the optimization isolation mechanism may prevent the additional shared capacity from being monopolized by a single task.
However, the improvements also exhibit clear diminishing returns. Therefore, we ultimately choose the layer of 3 and the hidden dimension of 640 for the online service to balance model performance and resource consumption.

\subsubsection{Inference Latency}

\begin{figure}[t!]
\begin{center}
\includegraphics[width=\linewidth]{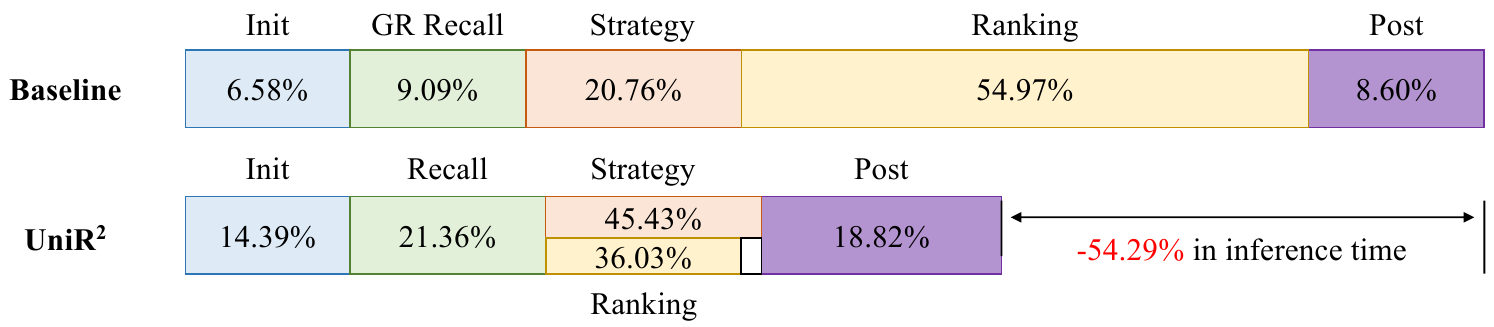}
\caption{The comparison of system inference time.}
\label{fig:time}
\end{center}
\end{figure}

We compare the end-to-end latency breakdown between the online pipeline and UniR$^2$, with the results shown in Figure \ref{fig:time}. 
In the baseline, recall and ranking are executed by two independent models, with the ranking stage dominating the overall inference time. Together with repeated initialization, feature preparation, and information transfer across pipeline, this leads to redundant time consumption.
UniR$^2$ instead reuses the same user context and decoder-only architecture for both tasks, and appends the feature fusion required for ranking after the generated SID trajectory. 
As a result, reusing cached representations substantially reduces the overall pipeline latency. 
In addition, our carefully designed parallelization strategy and ranking scheme further reduce end-to-end inference time.
This result shows that unification provides not only representation and objective consistency, but also a concrete serving advantage by eliminating redundant computation and hiding ranking latency within the existing pipeline schedule.

\subsection{Online A/B Test (RQ3)}

To evaluate the real-world business impact of UniR$^2$, we deploy it to the live-streaming services of both Kuaishou and Kuaishou Lite App. 
UniR$^2$ simultaneously replaces the original OneLive-based recall model and the deployed production pre-ranking model, which together serve as the baseline assembled in cascaded architecture.
The online A/B test was launched in June 2026, lasted for two weeks, and exposed to 5\% of online traffic. 

As shown by the online results, UniR$^2$ brings consistent gains on both platforms. 
On Kuaishou App, it improves play volume by \textbf{+1.177\%}, follow rate by \textbf{+0.655\%} and like rate by \textbf{+2.560\%}. 
On Kuaishou Lite App, it improves the number of gifting users by \textbf{+0.717\%}, gifting intention by \textbf{+1.567\%}, and total gifting amount by \textbf{+2.569\%}. 
The positive improvements across multiple behavior metrics demonstrate that UniR$^2$ can both retrieve suitable candidates and produce high-quality user-item pair scores within a single model. 
These online gains further validate the practicality and effectiveness of unified recall and ranking in large-scale industrial recommendation systems.

\section{Conclusion}

In this work, we propose UniR$^2$, a unified decoder-only Transformer that organizes user context, SID trajectory, and item features within a single heterogeneous sequence. Its Dual-Query Prefix-Causal Attention enables causal SID generation and expressive ranking feature fusion under task-specific visibility, while task-specific forward paths and ranking-side LoRA balance forward representation sharing with optimization isolation. 
Extensive offline experiments demonstrate consistent improvements in both generative recall and multi-objective ranking, and further verify that the SID trajectory serves as an effective representation bridge between the two tasks. 
By reusing cached user-context and SID states, UniR$^2$ also reduces end-to-end inference time. 
Long-term online A/B tests show consistent positive gains, validating its effectiveness and practicality in large-scale industrial recommendation systems.

\balance
\bibliographystyle{ACM-Reference-Format}
\bibliography{sample-base-extend.bib}

\end{document}